\documentclass[12pt,a4paper]{article}
\usepackage[scale={.75,.8}]{geometry}
\usepackage{pstricks}
\usepackage[latin2]{inputenc}
\usepackage{appendix}
\usepackage{graphicx}
\usepackage{amssymb}
\usepackage{amsmath}

\newcommand{\be}{\begin{equation}}
\newcommand{\ee}{\end{equation}}
\newcommand{\ba}{\begin{eqnarray}}
\newcommand{\ea}{\end{eqnarray}}
\newcommand{\tr}{{\rm Tr}}
\newcommand{\pl}{\left\{}
\newcommand{\pr}{\right\}}
\newcommand{\al}{\left|}
\newcommand{\ar}{\right|}
\newcommand{\rr}{\right)}
\newcommand{\rl}{\left(}
\newcommand{\ccl}{\left[}
\newcommand{\ccr}{\right]}
\newcommand{\dd}{\partial}
\newcommand{\ffitil}{\widetilde \varphi}
\newcommand{\ffi}{\varphi}
\newcommand{\Phib}{\overline{\Phi}}

\newcommand{\chib}{\bar \chi}
\newcommand{\Fb}{\overline F}

\newcommand{\Wb}{\overline W}
\newcommand{\Tb}{\overline T}

\def\unit{\relax{\rm 1\kern-.26em I}}
\def\Phat{\widehat \Phi}
\numberwithin{equation}{section}
\begin{document}
\setcounter{page}{0}

\vspace{-1truecm}

\rightline{DESY-08-012}

\vspace{3.cm}

\begin{center}

{\huge {\bf Finite temperature behaviour of}}

\vspace{0.8cm}

{\huge {\bf the ISS-uplifted KKLT model}}

\vspace{1.5 cm}

{\large {\bf Chlo\'e Papineau}}

\vspace{1.5cm}

Deutsches Elektronen-Synchrotron DESY, Notkestrasse 85, 22607 Hamburg, Germany
\vspace{0.8cm}\\

 \texttt{chloe.papineau@desy.de}
\vspace{0.8cm}\\

\end{center}

\vspace{1cm}

\abstract{We study the static phase structure of the ISS-KKLT model for moduli stabilisation and uplifting to a zero cosmological constant. Since the supersymmetry breaking sector and the moduli sector are only gravitationally coupled, we expect negligible quantum effects of the modulus upon the ISS sector, and the other way around. Under this assumption, we show that the ISS fields end up in the metastable vacua. The reason is not only that it is thermally favoured (second order phase transition) compared to the phase transition towards the supersymmetric vacua, but rather that the metastable vacua form before the supersymmetric ones. This nice feature is exclusively due to the presence of the KKLT sector. We also show that supergravity effects are negligible around the origin of the field space. Finally, we turn to the modulus sector and show that there is no destabilisation effect coming from the ISS sector.}

\newpage

\vspace{3cm}

\newpage

\tableofcontents

\vspace{3cm}


\pagestyle{plain}

\vspace{-1truecm}


\section{Introduction}


In the last years, quite a large attention has been given to the problem of moduli stabilisation, especially concerning the cosmological implications they could have \cite{Dine:1999tp}. Following an earlier proposal \cite{Giddings:2001yu}, Kachru, Kallosh, Linde and Trivedi (KKLT, \cite{Kachru:2003aw}) have recently provided the first explicit model in which all moduli are fixed. They do so by turning on fluxes in a first step, which fix the complex moduli and the dilaton $S$, and introducing non-perturbative superpotentials \cite{Taylor:1982bp} in a second step in order to stabilise the K\"ahler moduli $T$. For a more detailed study of the phenomenology arising from these models, see \cite{Choi:2004sx}.

Unfortunately, the resulting low energy potential for $T$ has an anti-de Sitter vacuum which needs to be uplifted. The strategy proposed in \cite{Kachru:2003aw} was to introduce an anti-$D3$ brane far away from the visible sector in the sense of the compact dimensions so that the fine-tuning of the cosmological constant would be naturally explained by their curvature. However, such a mechanism results in a non-linearly realised supersymmetry, and therefore the low energy theory can no longer be expressed in terms of usual 4D supergravity. Latter attempts tried to realise the uplift by $D$-terms \cite{Burgess:2003ic}, i.e. using a fully supersymmetric sector, but this generically leads to a heavy gravitino\footnote{Actually this is not the case for the last two references of \cite{Burgess:2003ic} because the uplift there is mainly realised by an $F$-term.}.

Parallel works have considered an $F$-term uplifting \cite{Saltman:2004sn}. This relies on adding a new sector in which supersymmetry is spontaneously broken by some field $\Phi$, $F_{\Phi} \neq 0$. If this sector and the KKLT setup are decoupled in such a way that the K\"ahler potential and superpotentials add up, then the uplifting is trivially realised by relating the parameters of both sectors. The last two years, a rather large sample of such models together with their direct phenomenology have been proposed \cite{Dudas:2006gr,Abe:2006xp,Kallosh:2006dv}.

In this paper, we shall focus on the setup developped in \cite{Dudas:2006gr}, where the uplifting sector was chosen to be the Intriligator, Seiberg and Shih model (ISS, \cite{Intriligator:2006dd}). A non-exhaustive list of string realizations of it can be found in Ref. \cite{Franco:2006es}. This dual SQCD model is of particular interest since it realises a breaking of supersymmetry in local minima in the squarks direction. Elsewhere in the field space, in the mesons direction, there are supersymmetric vacua, and both locally stable points are separated by a potential barrier. This ensures a long life-time for the SUSY breaking vacua. Hence, one does not have to give up the idea of a global (supersymmetric) minimum. Now, from cosmological considerations, we are led to wonder whether we ended up living in the metastable vacuum or not. And indeed, following the results of Refs. \cite{Craig:2006kx,Abel:2006cr,Fischler:2006xh,Anguelova:2007at}, we do. There, the authors showed that finite temperature corrections\footnote{For a review on finite temperature field theory, see \cite{Quiros:1999jp}.} to the ISS setup favour the fields to go in the metastable vacua rather than in the supersymmetric ones. In \cite{Abel:2006cr}, it was assumed that the fields start in the supersymmetric phase. Instead, the authors of \cite{Fischler:2006xh,Anguelova:2007at} assumed the starting point to be the origin of the field space, which is a minimum at high temperature\footnote{We will develop these points in the following Sections.}. This is because the origin of the field space contains the highest number of light degrees of freedom and hence maximises the entropy. We shall adopt the same attitude. In these various studies, it was found that the supersymmetric vacua form at a higher temperature than the metastable ones, but the origin is always a local minimum in the mesons direction. Therefore, the phase transition is first order towards the supersymmetric vacua. This is thermally disfavoured in comparison with the second order phase transition that occurs towards the non-supersymmetric vacua, even though the latter happens at a lower temperature.

In this paper, we complete the study done in \cite{Anguelova:2007at}. We work out the complete phase structure of the ISS-KKLT model. As will become clear in the text, the ISS fields do end up in the non-supersymmetric vacua, and the modulus, on the other hand, is not destabilised by thermal effects, as suggested in \cite{Buchmuller:2004xr}. We will show that in our case, the presence of the modulus sector modifies the thermally corrected ISS picture in such a way that the metastable vacua form first, and they remain the true vacua of the theory during a certain time. Later on, the supersymmetric vacua form, but the fields have long gone in the SUSY breaking ones. Not until an even lower temperature are the two vacua degenerated. From that moment on, the fields can tunnel down from the metastable vacua to the supersymmetric vacua.

Another feature that was pointed point out in \cite{Anguelova:2007at} is that the origin of the ISS field space may no longer be a minimum at high temperature when this sector is coupled to the KKLT sector. This is due to supergravity, and could have a non-trivial effect on the phase transition. We study in great detail this point and find that as expected, this displacement is very small.

However, let us emphasize that this study is still at the toy model level. We will not at all address cosmological problems such as the gravitino overproduction that usually happens when the supersymmetry breaking sector is in thermal equilibrium. Even though the present paper obviously aims at a more realistic application, we leave these investigations for future work.

The paper is organised as follows. Section \ref{themodel} reviews the zero temperature ISS-KKLT setup in order for the paper to be self-contained. We introduce the main tools of finite temperature effective potential in Section \ref{sectionFT}. In Section \ref{TheISSsector}, the relevant temperatures and phase transition of the ISS sector are derived assuming the rigid limit (zeroth order in supergravity expansion). In subsection \ref{criticaltemp}, we compute the critical temperature of the second order phase transition towards the would-be metastable vacua. We give an insight of how the supersymmetric minima form in subsection \ref{formationsusy}. We eventually compute the degeneracy temperature between the non-supersymmetric and the supersymmetric vacua in subsection \ref{phasetransition}. The rigid limit assumption of Section \ref{TheISSsector} is verified by working out the supergravity corrections around the origin in Section \ref{displacedorigin}. Section \ref{Modulussection} deals with the modulus sector. We show that the temperature corrections coming from the thermalised ISS sector do not destabilise the modulus. Finally, we conclude in Section \ref{conclusions} and draw the future directions that seem relevant to us.


\section{ISS-KKLT model \label{themodel}}


Let us start by recalling the KKLT construction for moduli stabilisation in the framework of type IIB string theory. In \cite{Kachru:2003aw}, the authors used non trivial background fluxes, i.e. non-zero vacuum expectation values for certain field strengths in the internal directions, in order to stabilise all complex structure moduli as well as the dilaton. However, the K\"ahler modulus $T$, which describes the fluctuations of the overall internal volume, cannot be stabilised in this manner. Non-perturbative effects such as gaugino condensation on $D7$ branes are used to generate an Affleck-Dine-Seiberg \cite{Taylor:1982bp} superpotential at an intermediate scale $\Lambda \ll M_P$. At low energy, the procedure results in the following setup
\be
K_1 \ = \ - 3 \ \ln \rl T + {\Tb}\rr \quad , \quad W_1 \ = \ W_0 \ + \ a e^{-bT} \quad ,
\label{KKLTsector}
\ee
where the constant $W_0$ is remnant of the stabilisation of all other moduli at the Planck scale.

The model exhibits a supersymmetric minimum $D_T W_1 = \dd_T W_1 + K_T W_1 = 0$ at $T = T_0$, implying
\begin{eqnarray}
&&W_0 \ = \ - \ a e^{-b T_0} \pl 1 \ + \ \frac{b \rl T_0 + \Tb_0 \rr}{3} \pr \ < \ 0 \quad , \label{eqnT0} \\
&&\langle V_{\rm KKLT} \rangle = \langle e^{K_1} \ccl K^{T \Tb} D_T W_1 D_{\Tb} \Wb_1 - 3 \al W_1 \ar^2 \ccr \rangle = - \ \frac{a^2 b^2 e^{-b \rl T_0 + \Tb_0 \rr}}{3 \rl T_0 + \Tb_0 \rr} \ < \ 0 \quad , \nonumber
\end{eqnarray}
where $K^{T \Tb}=\rl K^{-1}\rr_{T \Tb}$ is the inverse metric for the K\"ahler potential $K_1$.\\

As mentionned in the Introduction, the energy can be uplifted to a positive value by adding a sector in which supersymmetry is spontaneously broken. In \cite{Dudas:2006gr}, the uplifting sector was chosen to be the ISS model \cite{Intriligator:2006dd}
\be
K_2 \ = \  \tr  \al \ffi \ar^2 \ + \ \tr \al \ffitil \ar^2 \ + \  \tr  \al \Phi \ar^2 \quad , \quad W_2 \ = \  h \ \tr \rl \ffitil \Phi \ffi \rr - h \mu^2 \ \tr \, \Phi  \quad .
\label{ISSsector}
\ee
This is the magnetic dual of a SUSY-QCD theory with gauge group $SU(N_c)$. When the number of flavours satisfies $N_f \leqslant 3 N_c /2$, the electric theory is asymptotically free whereas its dual, with gauge group $SU(N_f - N_c)$, is infrared free.

The magnetic fields under consideration are the gauge singlets $\Phi = \rl \Phi^i_j \rr$, which we call mesons because they are in one-to-one correspondence with the electric mesons. The quarks $\ffi = \rl \ffi^i_a \rr$, and the anti-quarks $\ffitil = \rl \ffitil^a_i \rr$ are in the fundamental and antifundamental representations of $SU(N)$. In the rigid supersymmetry limit, the theory (\ref{ISSsector}) has a global symmetry $G = SU(N_f)_L \times SU(N_f)_R \times U(1)_B \times U(1)' \times U(1)_R$ which is explicitly broken to $SU(N_f) \times U(1)_B \times U(1)_R$ by the mass parameter $\mu$.

We denote by $N$ the magnetic number of colours $N = N_f - N_c$, which satisfies $N_f \geqslant 3N$. The indices run as $i,j =1,\ldots, N_f$ and $a=1,\ldots, N$. For convenience, we will omit the flavour and colour indices from here on and will just keep in mind that $\Phi$ is an $N_f \times N_f$ matrix, whereas $\ffi$ and $\ffitil^T$ are $N_f \times N$ matrices.

The setup (\ref{ISSsector}) has supersymmetry breaking solutions
\be
\ffi \ = \ \ffitil^T \ = \ \begin{pmatrix} \mu \unit_{N} \cr 0 \end{pmatrix} \quad , \quad \Phi = 0
\label{issvacua}
\ee
generated by non-vanishing $F$-terms for the mesons $F_{\Phi} = h \rl \ffitil \ffi - \mu^2 \unit_{N_f}\rr$. Notice that the supersymmetry breaking does not affect the gauge sector since it is driven by gauge singlets. The corresponding vacuum energy is
\be
V_{\rm min} = \al h^2 \mu^4 \ar \, \rl N_f - N \rr \quad .
\label{energyISS}
\ee

Far away from the origin in the mesons direction, after integrating out the quarks, gaugino condensation produces a non-perturbative superpotential \cite{Taylor:1982bp}
\be
W_{\rm dyn} \ = \ N \rl \frac{h^{N_f} \det \Phi}{\Lambda_m^{N_f-3N}} \rr \quad ,
\label{Wdynmagn}
\ee
which gives rise to  supersymmetric vacua
\be
\langle h \Phi \rangle = \Lambda_m \epsilon^{2N/(N_f-N)} \unit_{N_f} \quad .
\label{susyvacua}
\ee
In the above expressions, $\Lambda_m$ is the dynamical scale of the magnetic theory, and $\epsilon \equiv \mu / \Lambda_m$ is a small parameter. The existence of these vacua renders the non-supersymmetric ones (\ref{issvacua}) metastable. Both regions of the ISS field space are separated by a potential barrier. The lifetime of the metastable vacua can be made arbitrarily large by tuning $\epsilon$ very small, or equivalently, $\Lambda_m$ very large for $\mu$ fixed.\\

We now couple both sectors in the following way
\be
K \ = \ K_1 \rl T , \Tb \rr  \ + \ K_2 \rl \chi^i , \chib_{\bar j} \rr \quad , \quad W \ = \ W_1 \rl T \rr \ + \ W_2 \rl \chi^i \rr \quad ,
\label{ISSKKLT}
\ee
where $\chi^i$ denote collectively the ISS fields $\ffi, \ffitil, \Phi$.

As explained in \cite{Dudas:2006gr}, such a decoupling between the two sectors can be achieved by considering systems of $D3$ and $D7$ branes. The gauge sector $SU(N)$ arises from a stack of $N$ $D3$ branes. Therefore the ISS gauge coupling, the dynamical scale $\Lambda_m$ and the mass parameter $\mu$ depend on the dilaton, which was already stabilised at higher energies. The mesons are interpreted as the positions of $N_f$ $D7$ branes, this ensures the decoupling in the K\"ahler potential (\ref{ISSKKLT}). The (anti-)quarks, on the other hand, are seen as open strings in the $D3$-$D7$ sector. Thus, their kinetic terms may not be canonical, but modifying the K\"ahler potential does not affect the main picture of the model since they are not directly related to the supersymmetry breaking, and hence to the uplifting mechanism.

The supergravity corrections are negligible around the metastable vacua, as they are higher order terms in powers of the ISS fields $\sim \mu^2 / M_P^2$. There, the scalar potential is well approximated by
\be
V \rl \chi^i, \chib_{\bar i}, T, \Tb \rr \ \simeq \ \frac{1}{\rl T + \Tb \rr^3} \ V_{\rm ISS} \rl \chi^i, \chib_{\bar i}\rr \ + \ V_{\rm KKLT} \rl T, \Tb \rr \quad ,
\label{expansionpotential}
\ee
where $V_{\rm ISS}$ is the global supersymmetric (as opposed to supergravity) scalar potential for the ISS sector. However, when computing the critical temperature, we will consider the expansion (\ref{expansionpotential}) to be valid at the origin of the field space as well. We then explicitly verify it in Section \ref{displacedorigin}.

\begin{figure}[htbp]
\begin{center}
\hspace{-1.3cm}
\includegraphics[scale=0.7]{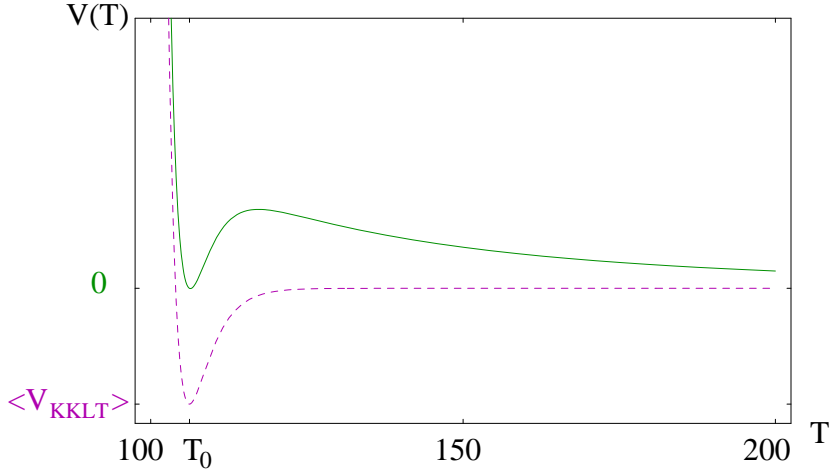}
\caption{The KKLT potential (purple, dashed) and the uplifted ISS-KKLT potential (green, plain). The vev $T \simeq T_0$ is not modified by the uplifting mechanism.\label{upliftplot}}
\end{center}
\end{figure}

The fine-tuning of the cosmological constant to zero is given by
\be
\langle V \rangle \  = \ 0 \quad \Longrightarrow \quad \al h^2 \mu^4 \ar \, \rl N_f - N \rr \ \simeq \ 3 \al W_0 \ar^2 \quad ,
\label{cosmconst}
\ee
and illustrated in Figure \ref{upliftplot}.

On the other hand, the gravitino mass is
\be
m_{3/2}^2 \ = \ \langle e^{K} \al W \ar^2 \rangle \ \simeq \ \frac{\al W_0 \ar^2}{\rl T_0 + \Tb_0 \rr^3} \ \simeq \ \frac{a^2 b^2 e^{- b \rl T_0 + \Tb_0 \rr}}{9 \rl T_0 + \Tb_0 \rr} \quad ,
\label{m3/2}
\ee
where (\ref{eqnT0}) together with the condition $b T_0 \gg 1$ were used in the last equality\footnote{As usual, we assume that the uplift of the modulus potential does not substantially modify its vev. This can be easily verified graphically (Figure \ref{upliftplot}).}.\\

Here and in the following numerical results, we fix $a = h =1$ and $b = 0.3$. We also set $N_f = 7$ and $N=2$. Asking for a TeV range gravitino mass and imposing (\ref{cosmconst}), the parameters are found to be
\be
T_0 \simeq 110 \quad , \quad \al W_0 \ar \simeq \rl 10^{-14} \, - \, 10^{-13} \rr \, M_P^3 \quad , \quad \mu \simeq 1\, - \,  5 \cdot 10^{-7} \, M_P \quad .
\label{parameters}
\ee
When needed, we will also take a coupling constant $g = 0.1$. Since the ISS gauge sector lives on $D3$ branes, one should have $g \rl M_P \rr \sim 4 \pi / {\rm Re} \,S$ and run this value down to a scale of order $\mu$. However, the dilaton vev strongly depends on the UV completion of the model. Though we believe this point to be crucial, it goes far beyond the aim of the present work. We also consider the scale $\Lambda_m \simeq 10^{11} M_P$. Let us recall that this scale is a Landau pole, which is not physical, and it is therefore not surprising to have $\Lambda_m$ higher than the Planck scale. The previous value corresponds to $g \rl M_P \rr = 0.5$ which means that the theory is perturbative at the Planck scale.

Before coming to the specificities of our model, let us introduce the finite temperature formalism.


\section{Finite temperature effective potential \label{sectionFT}}


The general one-loop effective potential including finite temperature effects can be split into different contributions \cite{Dolan:1973qd}
\be
V_{\rm eff}(\chi^i, T)  \ = \  V_{0}(\chi^i, T) \ + \ V_1^{0}(\chi^i, T) \ + \ V_1^{\Theta}(\chi^i, T) \quad ,
\label{potentialatFT}
\ee
where
\be
V_0 \ = \ e^K \pl K^{T \Tb} D_T W D_{\Tb} \Wb + K^{i \bar j} D_i W D_{\bar j} \Wb - 3 \al W \ar^2 \pr 
\label{potentialsugra}
\ee
is the tree-level supergravity potential, and $K^{i \bar j}$ is the inverse metric for the K\"ahler potential $K_2$.

The potential $V_1^0$ is the usual one-loop temperature independent Coleman-Weinberg effective potential \cite{Coleman:1973jx}, and $V_1^{\Theta}$ is the finite temperature contribution
\begin{eqnarray}
V_1^{\Theta}& = &\frac{\Theta^4}{2\pi^2}\pl \sum_B n_B \int_0^\infty dx \ x^2 \ln \rl 1-e^{-\sqrt{x^2 + M_B^2/\Theta^2}}\rr \right. \nonumber \\
&\quad& \quad \quad - \left. \sum_F n_F \int_0^\infty dx \ x^2 \ln \rl 1+e^{-\sqrt{x^2 + M_F^2/\Theta^2}}\rr \pr \quad .
\label{Vtheta}
\end{eqnarray}
Here $n_B$ ($n_F$) are the bosonic (fermionic) degrees of freedom, and $M_B$ ($M_F$) are the bosonic (fermionic) field-dependent mass eigenvalues. One can immediately see that finite temperature corrections break supersymmetry.

The potential (\ref{Vtheta}) may be expanded at high temperature, $\Theta^2 \gg M_B^2, M_F^2$,
\be
V_1^{\Theta} \simeq - \frac{\pi^2 \Theta^4}{90}\rl n_B + \frac{7}{8} n_F\rr \ + \ \frac{\Theta^2}{24} \, \rl \, 3 \tr \,\mathcal{M}_v^2 + \tr \,\mathcal{M}_{f}^2 + \tr \,\mathcal{M}_s^2 \, \rr  +  \ldots \quad ,
\label{VthetaHT}
\ee
where $\mathcal{M}_x$ are the mass matrices for vectors, fermions, and scalars, expressed in terms of the fields. The trace $\tr \, \mathcal{M}_f^2$ is summed over Weyl fermions.

In general, one should use the following supergravity formul\ae{} for the mass matrices in the presence of a non-canonical K\"ahler potential \cite{Binetruy:1984wy}
\be
\tr \, \mathcal{M}_f^2 = \langle  e^G \ccl K^{A \overline B} K^{C \overline D} \rl \nabla_A G_C + G_A G_C \rr \rl \nabla_{\overline B} G_{\overline D} + G_{\overline B} G_{\overline D} \rr - 2 \ccr \rangle \quad ,
\label{TrMfsugrageneral}
\ee
and
\be
\tr \, \mathcal{M}_s^2 = 2 \, \langle \, K^{A \overline B}  \frac{\dd^2 V_0}{\dd \chi^A \dd \chib^{\overline B}} \, \rangle \quad .
\label{TrMssugrageneral}
\ee
In the above expressions, $\chi^A$ represent the scalar fields in thermal equilibrium. The term $-2$ entering (\ref{TrMfsugrageneral}) takes into account the mixed Goldstino-gravitino contribution \cite{Binetruy:1984wy}. The function $G = K + \ln \al W \ar^2$ is the supergravity K\"ahler invariant potential, and we also introduced $G_A = \dd G / \dd \chi^A$ and $\nabla_A G_B = G_{AB} - \Gamma^C_{AB} G_C$, with the connection
\be
\Gamma^C_{AB} \ = \ K^{C \overline D} \dd_A K_{B \overline D} \quad .
\label{connection}
\ee

However, as briefly mentionned in the Introduction, we will be concerned with these general results in Section \ref{displacedorigin} when we explicitly calculate how supergravity together with temperature effects displace the minimum from the origin in the mesons direction, and in Section \ref{Modulussection} when we study the destabilisation of the modulus. Since both $T$ and $\Phi$ are singlets under $SU(N)$, the gauge bosons contribution will not be relevant when computing the derivatives of the effective potential (\ref{VthetaHT}). This is why we did not write $\tr \, \mathcal{M}_v^2$ here above.

At the origin, we keep the ISS sector at the rigid level. When computing the finite temperature corrections there, we shall use the results of global supersymmetry
\begin{eqnarray}
3 \, \tr \, \mathcal{M}_v^2  \ &=& \ 6 \, \langle \, D^{\alpha}_i D^{\alpha \, i} \, \rangle \quad , \nonumber \\
\tr \, \mathcal{M}_f^2 \ &=& \ \langle \, \Fb^{i j} F_{i j} \, \rangle \ + \ 4 \, \langle \, D^{\alpha}_i D^{\alpha \, i} \, \rangle \quad , \label{trM2susy} \\
\tr \, \mathcal{M}_s^2 \ &=& \ 2 \, \langle \, \Fb^{i j} F_{i j} \, \rangle \ + \ 2 \, \langle \, D^{\alpha}_i D^{\alpha \, i} \, \rangle \quad , \nonumber
\end{eqnarray}
where, as usual, $F_{i j} = \dd^2 W / \dd \chi^i \dd \chi^j$ and $D^{\alpha}_i = \dd D^{\alpha} / \dd \chi^i$. Here again, $\chi^i$ represent the scalar fields associated with $\ffi$, $\ffitil$ and $\Phi$. One should be aware that the traces above run over the flavour and colour indices as well. The index $\alpha$ labels the adjoint representation of $SU(N)$.

Let us now turn to the main part of this paper, namely the phase structure of the model (\ref{ISSKKLT}) once finite temperature corrections are included.


\section{Critical temperature and phase transitions in the ISS sector\label{TheISSsector}}


When the Universe cools down, the ISS fields end up in the non-supersymmetric vacua, as studied in \cite{Craig:2006kx,Abel:2006cr,Fischler:2006xh,Anguelova:2007at}. In this section, we show that the picture is not drastically modified when we add the modulus sector. However, this scheme is valid only if we consider the KKLT sector to be classical, which means that we assume the modulus to be already lying in its minimum $T = T_0$ and we neglect its quantum corrections to the ISS sector. In turn, Section \ref{Modulussection} deals with the eventuality of a modulus destabilisation by temperature.

Finite temperature effects are to restore all symmetries. At sufficiently high temperature, all the fields sit at the origin of the ISS field space. As we shall see, when the temperature lowers, the potential starts to exhibit a tachyonic direction towards the non-supersymmetric vacua, which form first. At a lower temperature, the would-be supersymmetric vacua form, but the origin remains a local minimum in the mesons direction (saddle point). Therefore, the origin and these new minima are separated by a barrier.

\subsection{Critical temperature \label{criticaltemp}}


In what follows, we focus on the behaviour of the potential at the origin of the field space. The symmetry restoration due to finite temperature appears when the tachyonic tree level masses are compensated by the thermal masses (second derivatives of the potential (\ref{Vtheta})) at the origin. This is also a good reason to keep the lowest order in supergravity around the origin. Indeed, even though the corrections to the potential are negligible, supergravity effects could have a non-trivial impact on its derivatives and one typically has to take them into account. However, in the case of the critical temperature, and motivated by the results of \cite{Anguelova:2007at}, even if the exact location of the origin may vary with temperature and supergravity, the moment when the curvature of the potential at the origin becomes negative should not be drastically affected by supergravity effects. This, obviously, assumes that the origin is indeed a minimum at high temperature, as we will show in Section \ref{displacedorigin}.

We use the high temperature expansion (\ref{VthetaHT}) because the tree level masses are of order $h^2 \mu^2 / \rl T_0 + \Tb_0 \rr^3$, see for instance (\ref{parameters}). We follow the standard procedure \cite{Dolan:1973qd} but there is no need to shift the fields here since we work at the origin of the field space.

The traces (\ref{trM2susy}) expressed in terms of the fields are easily calculated from the superpotential (\ref{ISSsector}). We find
\begin{eqnarray}
3\, \tr \mathcal{M}_v^2 &=& 3 g^2 \frac{N^2 - 1}{N}\  \frac{\tr \al \ffi \ar^2 + \tr \al \ffitil \ar^2}{\rl T_0 + \Tb_0 \rr^3}\quad , \label{trM2origin} \\
\tr \mathcal{M}_f^2 &=& 2 \rl h^2 N_f + g^2 \frac{N^2 -1}{N} \rr  \frac{\tr \al \ffi \ar^2 + \tr \al \ffitil \ar^2}{\rl T_0 + \Tb_0 \rr^3} + 2h^2 N \frac{\tr \al \Phi \ar^2}{\rl T_0 + \Tb_0 \rr^3} \ , \nonumber \\
\tr \mathcal{M}_s^2 &=& \rl 4 h^2 N_f + g^2 \frac{N^2 -1}{N} \rr \frac{\tr \al \ffi \ar^2 + \tr \al \ffitil \ar^2}{\rl T_0 + \Tb_0 \rr^3} + 4h^2 N \frac{\tr \al \Phi \ar^2}{\rl T_0 + \Tb_0 \rr^3} \ , \nonumber
\end{eqnarray}
where $g$ is the coupling constant. It follows that the potential (\ref{VthetaHT}) reads
\be
V_1^{\Theta} =  \frac{\Theta^2}{4 \rl  T_0 + \Tb_0  \rr^3} \pl \rl h^2 N_f + g^2 \frac{N^2 -1}{N} \rr \ccl \tr \al \ffi \ar^2 + \tr \al \ffitil \ar^2 \ccr + h^2 N \tr \al \Phi \ar^2\pr \ , \label{V1thetaHTorigin}
\ee
where we dropped the constant term $\propto \Theta^4$ in (\ref{VthetaHT}) since it is not relevant for the computation of the critical temperature.

We compare the scalar thermal masses in (\ref{V1thetaHTorigin}) to the tree level masses at the origin. The latter are $\pm \, h^2 \mu^2 / \rl T_0 + \Tb_0 \rr^3$ or $\pm \, h^2 \mu^{*\, 2} / \rl T_0 + \Tb_0 \rr^3$ for the squarks and $0$ for the mesons, as easily seen from the superpotential (\ref{ISSsector}) and from the expansion (\ref{expansionpotential}). The thermal mass matrix is diagonal and positive definite, while the classical mass matrix is anti-diagonal. We ask for the determinant of the whole squared mass matrix to be zero at the critical temperature $\Theta_c$. This means that all the eigenvalues are positive above the critical temperature, while below $\Theta_c$, tachyonic directions appear in the potential towards the would-be metastable vacua. From (\ref{V1thetaHTorigin}), we get
\be
\Theta_c^2 \ = \ \frac{4 \al \mu^2 \ar}{N_f + \frac{g^2}{h^2} \frac{N^2-1}{N}} \quad , \label{ThetaC}
\ee
which is in agreement with the critical temperature derived in \cite{Anguelova:2007at} in the rigid limit. As computed there, this is only slightly modified by supergravity corrections.

The critical temperature (\ref{ThetaC}) is of order $\mu^2$ and the tree-level masses are of order $h \mu^2 / \rl  T_0 + \Tb_0  \rr^3$, with $T_0 \sim 110$. Therefore, the use of the high temperature expansion (\ref{VthetaHT}) is a posteriori justified.

For the values of the parameters given in (\ref{parameters}), i.e. for a gravitino mass $m_{3/2} = 1$ TeV, we find that $\Theta_c \simeq 4 \cdot 10^{-7} M_P$. Notice that the critical temperature does not depend on the modulus, as one could have expected from the expansion (\ref{expansionpotential}). This is a crucial point since it is the reason why the would-be metastable vacua form first, as we show now.

\subsection{Formation of the supersymmetric vacua\label{formationsusy}}


Having computed the critical temperature does not yet ensure that the ISS fields actually go in the metastable vacua. In this section, we turn to the mesons direction and work out the temperature $\Theta_{\rm susy}$ at which the SUSY preserving vacua appear. In particular, we want to know if they are already formed when $\Theta = \Theta_c \,$. In order to achieve this, we ask for the mesons to be away from the origin and integrate out the heavy quarks. The low energy theory is then pure Yang-Mills, it is strongly coupled in the IR and gaugino condensation \cite{Taylor:1982bp} produces the non-perturbative term (\ref{Wdynmagn}) which gives rise to the supersymmetric vacua
\be
W_{\rm NP} \ = \ N A \left( \det \Phi \right)^{1/N} \ ,
\label{WNP}
\ee
with $A = h^{\nu} \Lambda_m^{-\nu + 3}$  and $\nu = N_f/N$.

At zero temperature, the vacua are $\Phat_0 = \rl A^{-1} h \mu^2 \rr^{1/(\nu -1)} \unit_{N_f}$ and the quarks masses are $m_{\ffi , \, \ffitil} = h\Phat_0 / \rl T_0 + \Tb_0 \rr^{3/2}$.

We decompose the mesons into a classical background $\Phat$ and a quantum field $\phi$ as follows
\be
 \Phi \ = \ \Phat \, \unit_{N_f} \ + \ \phi \quad .
\label{mesondecomp}
\ee

Expanding the total superpotential $h\tr \rl \ffitil \Phi \ffi \rr - h \mu^2 \tr \, \Phi + W_{\rm NP}$ according to the above decomposition, we get
\begin{eqnarray}
W &=& \rl N A \Phat^{\nu-1} - h\mu^2 N_f \rr \Phat + \rl A \Phat^{\nu-1} - h\mu^2 \rr \tr \phi + h \Phat   \tr \rl \ffitil \ffi \rr  \nonumber \\
&+&  h \tr \rl \ffitil \phi \ffi \rr + \frac{1}{2} \, A \Phat^{\nu - 2}\pl \frac{ \rl \tr \phi \rr^2}{N} - \tr \phi^2  \pr \ .
\label{Wmesons}
\end{eqnarray}
Keeping the quadratic order in $\phi$ is sufficient because, following the standard procedure \cite{Dolan:1973qd}, we express the masses in terms of the classical field $\Phat$ and hence higher powers in $\phi$ are not relevant. Notice also that the quarks $\ffi$ and $\ffitil$ should not be present in $W$ since they have been integrated out. This point will become clear as we advance in the computation.

At this stage, we would like to emphasize that working out the whole finite temperature corrected potential in the context of supergravity drives a lot of technical complications. For the sake of clarity, in order to sketch the mechanism that happens around the SUSY vacua, we will again consider the rigid limit. We believe that supergravity corrections do not strongly modify the following results.

If the quarks are integrated out, it means that the temperatures we consider are $\ \Theta \ll h \Phat_0 / \rl T_0 + \Tb_0 \rr^{3/2}$. On the other hand, using (\ref{trM2susy}), we find that the masses of the mesons are
\be
\tr \, \mathcal{M}_f^2 + \tr \, \mathcal{M}_s^2 \ = \ 3 \rl N_f^2 - 2 \nu + \nu^2 \rr \, \frac{A^2 \Phat^{2 \nu - 4}}{\rl T_0 + \Tb_0 \rr^3} \quad ,
\label{massmesons}
\ee
which is in agreement with \cite{Fischler:2006xh}. The high temperature expansion (\ref{VthetaHT}) is thus legitimate and one finds that the effective potential is
\be
V_1^{\Theta} \ = \ - C \Theta^4  \ + \ \frac{\rl N_f^2 - 2 \nu + \nu^2 \rr}{8} \, \frac{A^2 \Phat^{2 \nu - 4} \Theta^2}{\rl T_0 + \Tb_0 \rr^3}  \quad .
\label{VthetaMesons}
\ee
From this expression, we see that the thermal contribution to the mesonic mass matrix, namely the second derivative of (\ref{VthetaMesons}) with respect to $\Phat$, is diagonal and positive definite. Therefore there is no way that this contribution can lead to a destruction of the SUSY vacua and hence to a ``critical'' temperature. In other words, the masses are already positive at zero temperature and thus the origin and the supersymmetric vacua are separated by a barrier. When temperature effects are included, only a first order phase transition can happen.

Moreover, the contribution (\ref{VthetaMesons}) is much smaller than that of the tree-level masses $m_{\phi} \sim A \Phat^{\nu - 2} / \rl T_0 + \Tb_0 \rr^{3/2}$ by assumption\footnote{Since $T_0 \simeq 110$, the assumption $\Theta \ll h \Phat / \rl T_0 + \Tb_0 \rr^{3/2}$ implies that $\Theta$ is even smaller than $h \Phat$.}. Even if we used a high temperature expansion, recall that the thermal mass is proportional to the quartic self-coupling of the mesons. Since it arises from non-perturbative effects, it is unnaturally small.\\

Even though the squarks have been integrated out at tree level, their effect in the loops may be important. However, due to their large masses, the high temperature expansion can not be used. From (\ref{Vtheta}), we derive a low temperature expansion
\be
V_1^{\Theta} \ = \ - \ \frac{\Theta^{5/2}}{2 \pi^{3/2}} \, \pl \, \sum_B n_B M_B^{3/2} e^{- M_B / \Theta} \ + \ \sum_F n_F M_F^{3/2} e^{- M_F / \Theta} \,  \pr \quad ,
\label{VthetaLT}
\ee
with $M_{B , \, F} \gg \Theta$.

At the leading order, the squarks mass matrix is almost diagonal and its eigenvalues are $h^2 \Phat^2 / \rl T_0 + \Tb_0 \rr^3$. Recall this sector is supersymmetric, so the fermionic and bosonic degrees of freedom give the same contribution. Using (\ref{VthetaLT}), we find
\be
V_1^{\Theta}  \ = \  - \ \frac{3 N N_f}{\pi^{3/2}} \ \rl \frac{h \Phat}{\rl T_0 + \Tb_0 \rr^{3/2}} \rr^{3/2} \, \Theta^{5/2} \ \exp \ \pl - \, \frac{h \Phat}{\Theta \rl T_0 + \Tb_0 \rr^{3/2}} \pr \quad .
\label{VthetaQuarks}
\ee
This consists of a negative contribution which, together with the tree level potential deduced from (\ref{Wmesons}) and with the effective potential (\ref{VthetaMesons}), gives the total potential. One then has to consider the system
\be
\dd V_{\rm tot} / \dd \Phat \ = \ 0 \ = \ \dd^2 V_{\rm tot} / \dd \Phat^2 \quad . \nonumber
\ee
Solving it brings us to knowing the temperature $\Theta_{\rm susy}$ and the corresponding vev $\Phat \rl \Theta_{\rm susy} \rr$. However, it turns out to be very hard to solve and we approximate $\Phat = \Phat_0$ at all temperatures. We concentrate on the second equation of the expression above and find
\be
\Theta_{\rm susy} \ = \  \frac{h \Phat_0}{B \, \rl T_0 + \Tb_0 \rr^{3/2}}  \quad ,
\label{ThetaSusy}
\ee
with
\be
B = - \ln \ccl \frac{\rl N_f^2 - 2 \nu + \nu^2 \rr \pi^{3/2}}{h^4 N N_f^2} \, A^2 \Phat_0^{2\nu - 6} \rl T_0 + \Tb_0 \rr^{3} \ccr \ > \ 0 \quad . \nonumber
\ee
For $B \gg 1$, we fulfill the consistency condition that the squarks are integrated out at tree level.


\subsubsection*{Numerical results}


As we already explained at the end of Section \ref{themodel}, the dynamical scale $\Lambda_m$ of the theory relies on the UV completion of our model. Since it is not the object of this work, we choose to consider the case of a half-unit gauge coupling at the Planck scale. Then $\Lambda_m = M_P e^{- 2 \pi / (3 N - N_f) g^2 \rl M_P \rr}$ is approximately $10^{11} M_P$. Recall that $N_f \geqslant 3 N$ so that the argument of the exponential is positive.

For this value and the rest of the parameters given by (\ref{parameters}), we find the following results
\be
\Phat_0 \ \simeq \ 1 \cdot 10^{-3} \quad , \quad m_{\ffi , \, \ffitil} \ \simeq \ 4 \cdot 10^{-7} \quad , \quad m_{\phi} \ \simeq \ 3 \cdot 10^{-13}
\label{numericalresults}
\ee
for the vev of the mesons and for the tree level masses, and
\be
B \ \simeq \ 15 \quad \quad , \quad \quad \Theta_{\rm susy} \ \simeq \ 3 \cdot 10^{-8} \nonumber
\ee
for the temperature. All these results, except for $B$, are expressed in units of $M_P$. Notice that $B$ is larger than one.

The main conclusion is that the SUSY vacua form at a temperature which is smaller than the critical temperature (\ref{ThetaC}). Obviously, this result depends on the choice of $\Lambda_m$ and we emphasize, again, that a closer study of the UV physics of our model is required. However, we find numerically that the constant $B$ is negative for $\Lambda_m$ smaller than $10^4$ which corresponds to a gauge coupling of $0.9$ at the Planck scale. Therefore it seems that the more the theory is perturbative at high energy, the more consistent the picture is. Varying $\Lambda_m$ in this range yields
\be
10^4 \ \leqslant \ \Lambda_m \ \leqslant \ 10^{15} \quad \Longrightarrow \quad 1.5 \cdot 10^{-8} \ \leqslant \ m_{\ffi , \, \ffitil} \ \leqslant \ 7.4 \cdot 10^{-6} \quad . \label{rangeSUSYvacua}
\ee
Hence, even for very high values of $\Lambda_m$, the upper bound under which one can integrate out the squarks is only slightly above the critical temperature (\ref{ThetaC}), and the corresponding SUSY temperature is $3 \cdot 10^{-7} M_P \, \lesssim \Theta_c \,$.

It is clear that the major cause of such an effect is the explicit dependence of the SUSY temperature on the modulus. This pushes the tree level squarks masses to very low values compared to the original ISS scenario.

Another result that we were able to derive numerically is that already once the squarks are integrated out, their contribution (\ref{VthetaQuarks}) is very small compared to the tree-level one (\ref{Wmesons}). This means that whenever one can consider the non-perturbatively generated superpotential (\ref{WNP}), then the vacua are already there. As such, the SUSY temperature (\ref{ThetaSusy}) does not really make sense, and we are more encline to rely on the evaluation of the squarks masses $m_{\ffi, \, \ffitil}$ as in (\ref{rangeSUSYvacua}). Also, since these are tree level masses, they do not depend on $\Phat \rl \Theta \rr$ and thus are not biased by our approximations.

The conclusion is unchanged : the supersymmetric vacua form after the would-be metastable ones, and this is due to the presence of the modulus.

\subsection{Degeneracy between the vacua\label{phasetransition}}


Finally, in this paragraph we compute the degeneracy temperature $\Theta_{\rm deg}$ at which it becomes possible for the fields to go from the metastable vacua to the supersymmetric ones. This temperature is defined as the moment when both vacua have the same energy.

The total number of degrees of freedom in the non-supersymmetric vacua is $\rl N_f + N \rr^2 - 1$. The vacuum energy there is
\be
\left. \langle V \rangle \ar_{\rm meta} \ = \ - \frac{\pi^2 \Theta^4}{24} \ccl \rl N_f + N \rr^2 - 1 \ccr + \frac{\al h^2\mu^4 \ar \rl N_f - N \rr}{\rl T_0 + \Tb_0 \rr^3} \quad , \nonumber
\ee
where we did not account for the KKLT energy since it is constant over the whole ISS field space.

In the last paragraph, we showed that the squarks can be totally neglected in the supersymmetric vacua. Therefore only the finite (high) temperature correction coming from the mesons is relevant. Recalling that these vacua have zero energy at tree-level, using (\ref{VthetaMesons}) and assuming again that $\Phat = \Phat_0$, one finds
\be
\left. \langle V \rangle \ar_{\rm susy} \ = \ - \frac{\pi^2 \Theta^4}{24}  N_f^2 + \frac{\rl N_f^2 - 2 \nu + \nu^2 \rr}{8} \, \frac{A^2 \Phat^{2 \nu - 4} \Theta^2}{\rl T_0 + \Tb_0 \rr^3} \quad . \nonumber
\ee
Actually, it is easily seen from our numerical results (\ref{numericalresults}) that the last term in the above expression is negligible. To good approximation, the degeneracy temperature is thus given by\footnote{This result is in slight disagreement with \cite{Fischler:2006xh}. First of all, they computed the degeneracy temperature from the origin to the supersymmetric vacua. Indeed, when the KKLT sector is not present, the latter form before the metastable vacua. However, by dropping the $T$-dependence and replacing $N_f$ by $N_f - N$ in the prefactor of (\ref{ThetaDeg}), we do not find exactly their result. This is due to the fact that they use a high temperature expansion even for the squarks in the supersymmetric vacua, which results in dropping the $2N N_f$ in the denominator.}
\be
\Theta_{\rm deg}^2 \ \simeq \ \sqrt{\frac{24 N_f}{\pi^2 \rl 2 N N_f + N^2 - 1 \rr}}\  \frac{\al h \mu^2 \ar}{(T+{\bar T})^{3/2}} \quad .
\label{ThetaDeg}
\ee

Using our parameters, one finds numerically $\Theta_{\rm deg} \simeq 7 \cdot 10^{-9} M_P$. As before, the degeneracy temperature explicitly depends on the modulus, reason why it is so low compared to the critical temperature. We believe that this is a major improvement over the case of an isolated ISS sector. As was already noted in \cite{Dudas:2006gr}, the presence of the modulus enhances the lifetime of the ISS metastable vacua. We confirm this result here by showing that the supersymmetric vacua actually become the true vacua of the theory only at relatively late times.


\section{Supergravity and finite temperature corrections at the origin\label{displacedorigin}}


The computations of Section \ref{TheISSsector} have assumed that the origin of the ISS field space is a minimum of the potential at high temperature. However, as pointed out in \cite{Anguelova:2007at}, this is not as straightforward once one includes supergravity. Consider for instance the cross term $K^{T \Tb} K_T W_2 \dd_{\Tb} \Wb_1$ in the supergravity potential (\ref{potentialsugra}). It contains a linear term in $\Phi$ which contributes as a constant to the equation $\partial_{\Phi} V = 0$, and produces a displacement from the origin. Generical temperature corrections contain similar terms and one has to work out the full supergravity plus temperature corrected potential and solve for a minimum around the origin. This is an important point because, even though unexpected, the displacement could be large enough to spoil the phase transition towards the supersymmetry breaking vacua.

From the superpotential and K\"ahler potential (\ref{ISSsector}), one can see that only terms of at least quadratic order $\sim \ffi^2$, $\ffitil^2$, $\ffi \ffitil$ can appear in the scalar potential. Consequently, the origin $\ffi = \ffitil = 0$ is always a solution to the extremum equations $\dd_{\ffi} V = 0 = \dd_{\ffitil} V$. In what follows, we concentrate on the equation $\dd_{\Phi} V_{\rm eff} = 0$ in the background $\ffi = \ffitil =0$ (here $V_{\rm eff}$ stands for the full potential defined in (\ref{potentialatFT})).

The tree-level scalar potential
\be
V_0 \ = \ e^{K} \ccl K^{T \Tb} D_T W D_{\Tb} \Wb + K^{i \bar j} D_i W D_{\bar j} \Wb - 3 \al W \ar^2 \ccr 
\label{V0sugra}
\ee
receives temperature corrections given by (\ref{VthetaHT}), where the mass matrices squared (\ref{TrMfsugrageneral}) and (\ref{TrMssugrageneral}) can be developped using the semi-canonical K\"ahler potential (\ref{ISSKKLT})
\be
\tr \,\mathcal{M}_{f}^2 = \langle \, e^{G} \ccl  K^{i \bar k} K^{j \bar l} \rl G_{ij} + G_i G_j \rr  \rl G_{\bar k \bar l} + G_{\bar k} G_{\bar l} \rr  - 2 \ccr \, \rangle \quad ,
\label{TrMfsugra}
\ee
and
\be
\tr \,\mathcal{M}_{s}^2 \ = \ \langle \, 2 K^{i \bar j} \frac{\dd^2 V_0}{\dd \chi^i \dd \chib^{\bar j}} \, \rangle \quad .
\label{TrMssugra}
\ee

The new minimum at high temperature satisfies
\be
\left . \frac{\dd V_0}{\dd \Phi} + \frac{\Theta^2}{24}\, \frac{\dd}{\dd \Phi} \, \pl \tr \,\mathcal{M}_{f}^2 + \tr \,\mathcal{M}_{s}^2 \pr \ \ar_{\ffi = \ffitil = 0} \ = \ 0 \quad .
\label{dv/dphitot} 
\ee

Let us start with the zero-temperature potential (\ref{V0sugra}). Differentiating with respect to $\Phi$ yields
\begin{eqnarray}
\dd_{\Phi} V_0 \ &=& \ e^K \ccl K^{T \Tb} \pl \rl D_T W K_{\Phi} + W_{\Phi} K_T\rr \Wb_{\Tb} + K_{\Phi} W_T K_{\Tb} \Wb \pr \right. \nonumber \\ 
&\quad& \quad \quad + \left.  K^{i \bar j} D_{\bar j} \Wb \rl K_{\Phi} D_i W + K_i W_{\Phi} + W_{i \Phi}\rr + \Wb D_{\Phi} W \ccr \quad . \nonumber
\end{eqnarray}

Since we expect the displacement $\langle \Phi \rangle$ to be small, it is sufficient to keep the linear order in $\Phi$. One gets
\begin{eqnarray}
\langle \dd_{\Phi} V_0 \rangle &=& e^{K_1}\ccl  \Phib \pl K^{T \Tb} \rl D_T W_1 \Wb_{\Tb} + W_T K_{\Tb} \Wb_1 \rr + \al h^2 \mu^4 \ar N_f + \al W_1 \ar^2  \pr \right . \nonumber \\
&\quad& \quad \quad - \left. h \mu^2 \unit_{N_f} \pl K^{T \Tb} K_T \Wb_{\Tb} + \Wb_1 - 2 h \mu^{* \, 2} \, \tr \, \Phib \pr \ccr \quad ,
\label{dphiV0}
\end{eqnarray}
where $K_1$ and $W_1$ are the pure KKLT potentials defined in (\ref{KKLTsector}).

It is a long but straightforward computation to derive the other two contributions in (\ref{dv/dphitot}) ; some steps are given in the Appendix \ref{appendixorigin} for the interested reader.

The general solution to the linearised equation (\ref{dv/dphitot}) is of the form
\be
\Phi \rl T, \Tb, \Theta \rr \ = \ h \mu^2 \rl \frac{A + B \Theta^2}{C + D \Theta^2} \rr \cdot \unit_{N_f} \quad ,
\label{Phigeneral}
\ee
where $A, B, C, D$ are functions of $T$ and $\Tb$ only, and given in (\ref{functionsABCD}).

\begin{figure}[htbp]
\begin{center}
\hspace{-1.3cm}
\includegraphics[scale=0.6]{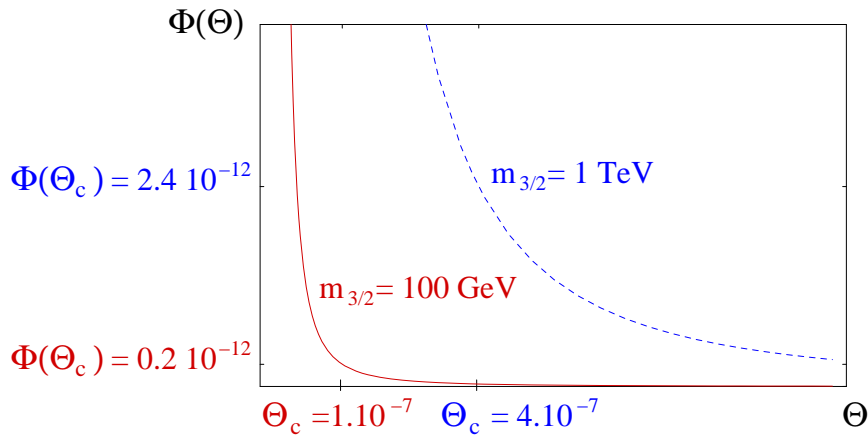}
\caption{Evolution of the mesons vev with the temperature for two values of the gravitino mass. When the temperature hits its critical value, the minimum at $\Phi$ turns into a saddle point.\label{mesonsplot}}
\end{center}
\end{figure}

Figure \ref{mesonsplot} shows the behaviour of (\ref{Phigeneral}) with respect to temperature for $T = T_0$. It is of some relevance to consider two different situations. For instance, the gravitino mass (\ref{m3/2}) fixes all the parameters, since the relation (\ref{eqnT0}) between $T_0$ and $W_0$ on the one hand, and the zero cosmological constant (\ref{cosmconst}) on the other hand are conditions of our model.

We choose to consider $m_{3/2} = 1$ TeV (blue, dashed line) and $m_{3/2} = 100$ GeV (red, plain line) as an example. In both cases, as expected, the origin is the only vacuum at very high temperature. One can already approximate $\Phi \sim 0$ at $\Theta \sim 10^{-6} M_P$ for the light gravitino case. The surprise comes from the fact that the minimum fades away from the origin very fast when the temperature lowers down, and this happens while the high temperature expansion is still valid. However it could very well be that keeping the linear order in $\Phi$ is no longer a good approximation there.

The critical temperature (\ref{ThetaC}) developped in Paragraph \ref{criticaltemp} depends on $\mu$ and thus on the gravitino mass. From Figure \ref{mesonsplot}, it is clear that the mesons still have a very small value at the critical temperature, for both cases we considered. Therefore, the phase transition towards the non-supersymmetric vacua, in the squarks direction, will not be affected by the displacement. A problem would have arised if the mesons vev had been too high (at the critical temperature), forcing us to take into account the non-perturbative superpotential (\ref{WNP}).

Moreover, the system with a light gravitino remains around the origin during a longer time, ensuring even more the phase transition. Indeed, one could find a set of parameters matching our two conditions (existence of a minimum for $T$, zero cosmological constant) for any gravitino mass. In the case of a substantially heavier gravitino, not only the volume modulus would have a too small vev, but the phase transition towards the would-be metastable vacua would be spoiled. We conclude that, even though it is not a very strong effect, our model seems to prefer a light gravitino.


\section{The modulus sector \label{Modulussection}}


Up to now, we have been considering that the modulus $T$ was sitting in its minimum. In this section, we shall derive the condition under which this is valid at the typical ISS temperatures.

Let us recall that since the moduli are only gravitationally coupled to the thermal bath, their interaction rate is
\be
\Gamma \ \simeq \ \frac{\Theta^3}{M_P^2} \quad \ll \quad H \ \simeq \ \frac{\Theta^2}{M_P} \quad . \nonumber
\ee
As such, the moduli potential is not in thermal equilibrium. However, indirect temperature corrections coming from other sectors could destabilise a modulus because they would result in an extra source of uplifting.

For instance, in \cite{Buchmuller:2004xr}, the authors studied the maximal (or critical) temperature beyond which a minimum generated by non-perturbative effects would be destroyed. Assuming that the visible sector lives on $D3$-branes, the gauge coupling is directly related to the vev of the dilaton $g^2 \sim 1/ \text{Re} \, S$. This implies that the dilaton potential is thermally perturbed through the gauge coupling. Typically, these effects destroy the minimum if they compensate the barrier between the metastable vacuum at $\text{Re} \, S \sim 2$ and the minimum at infinity. Therefore, in \cite{Buchmuller:2004xr}, the dilaton was destabilised for temperatures $\Theta \gtrsim \sqrt{m_{3/2} M_P} \simeq 10^{-8} M_P$ for a gravitino in the TeV range. The same could happen to the $T$-modulus if the visible sector lives on $D7$-branes.

In a KKLT setup, however, the dilaton is not stabilised by non-perturbative effects $W \sim e^{-b S}$, but rather by non-trivial background fluxes $W \sim m + n S$. Whereas gaugino condensation takes place at a scale $\Lambda \ll M_P$, resulting in a low mass for the modulus ($T$ or $S$ according to the model), a stabilisation by fluxes happens at high energy $\sim M_P$. The dilaton is then heavy enough not to be affected by temperature, and we can simply decouple it at low energy, as in the zero temperature theory. In what follows, we assume that this is the case, i.e. that the visible sector does live on $D3$-branes.

In our model, the $T$ modulus potential receives temperature corrections from the ISS sector. If there exists a critical temperature above which the potential is destabilised, we assume it to be higher than the temperatures computed in Section \ref{TheISSsector}. In this case, the ISS fields are at the origin, with the mesons slightly displaced, eq. (\ref{Phigeneral}).

We define the destabilisation temperature $\Theta_d$ and the corresponding value $T_d$ for the modulus as the point where the minimum turns into a saddle point :
\be
\frac{\dd V_{\rm eff}}{\dd T} \, \rl T_{d} , \Theta_{d} \rr \ = \ 0 \ = \ \frac{\dd^2 V_{\rm eff}}{\dd T^2} \, \rl T_{d} , \Theta_{d} \rr \quad ,
\label{eqnDestab}
\ee
where $V_{\rm eff}$ is the effective potential (\ref{potentialatFT}).

The computation follows similar steps as in Section \ref{displacedorigin} and Appendix \ref{appendixorigin}. We simply give here the result for the effective potential at linear order in the mesons displacement\footnote{Since we expect the destabilisation temperature to be very high, the linear approximation made in Section \ref{displacedorigin} is even more valid as one can convince oneself from Figure \ref{mesonsplot}.}
\be
V_0 \ = \ V_{\rm KKLT} \ + \ e^{K_1} \ccl \al h^2 \mu^4 \ar N_f - h \mu^2 A \tr \, \Phi - h \mu^{* \, 2} A^* \tr \, \Phib \, \ccr \nonumber \quad ,
\ee
where $A \rl T , \Tb \rr = K^{T \, \Tb} K_T \Wb_{\Tb} + \Wb_1$ was also defined in the Appendix \ref{appendixorigin}.

The traces of the mass matrices (\ref{TrMfsugra}) and (\ref{TrMssugra}) are
\be
\tr \,\mathcal{M}_{f}^2 = - 2 e^{K_1} \ccl \al W_1 \ar^2 - h \mu^2 \Wb_1 \tr \, \Phi - h \mu^{* \, 2} W_1 \tr \, \Phib \ccr \quad , \nonumber
\ee
and
\ba
\tr \,\mathcal{M}_{s}^2 &=& 2 e^{K_1} \ccl 2 N_f \rl N_f + 2N \rr \pl K^{T \Tb} D_T W_1 D_{\Tb} \Wb_1 - 2 \al W_1 \ar^2 \pr \right. \nonumber \\
&\quad& \quad \quad \quad + \pl 2 + 2 N_f \rl N_f + 2N \rr \pr \al h^2 \mu^4 \ar N_f \nonumber \\
&\quad& \quad \quad \quad - \left. \rl h \mu^2 \tr \, \Phi \pl 2 N_f \rl N_f + 2N \rr + 1 \pr \rl  K^{T \Tb} K_T \Wb_{\Tb} + 2 \Wb_1 \rr \, + {\rm h.c.} \rr \ccr \nonumber \quad .
\ea

This expression is easily implemented in a Mathematica routine in order to solve the system (\ref{eqnDestab}).

\begin{figure}[ht!]
\begin{center}
\includegraphics[scale=0.7]{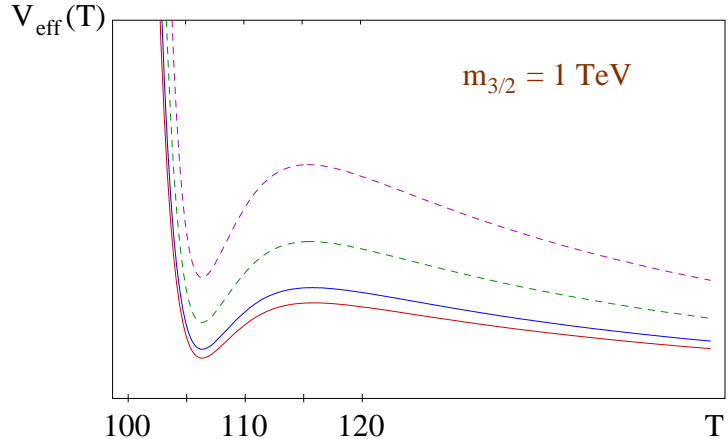}
\caption{Non-destabilisation of the $T$ modulus at high temperature. \label{modulusplot}}
\end{center}
\end{figure}

As a result, the one-loop effective potential is shown in Figure \ref{modulusplot}. We point out that the constant term $-C \Theta^4$ in (\ref{VthetaHT}) has not been included for graphical convenience. One can see that there is no destabilisation of the modulus at all, and indeed the system (\ref{eqnDestab}) turns out to be non-solvable.

We already argued in Section \ref{displacedorigin} that the parameters are fixed by the gravitino mass. In Figure \ref{modulusplot}, we took $m_{3/2} = 1$ TeV ; the rest of the parameters is the same, namely $a=1$, $h=1$, and $b=0.3$. The constants $W_0$ and $\mu$ are fixed by equations (\ref{eqnT0}) and (\ref{cosmconst}), and the solution $\Phi \rl T, \Theta \rr$ was derived in the Appendix \ref{appendixorigin}. This computation assumes that the value of $T$ at the minimum does not vary too much, which is cross-checked on Figure \ref{modulusplot}.

In \cite{Anguelova:2007ex}, the authors worked out the phase structure of the O'KKLT model \cite{Kallosh:2006dv}, which can be viewed as a simplified version of our model. Although they assumed the modulus to be in thermal equilibrium, it was found there that it is not destabilised by thermal corrections. In this perspective, we recover their result as the limit in which the thermal contribution of $T$ is negligible, which is indeed the case of interest for an expanding Universe.


\section{Conclusions and future challenges\label{conclusions}}


Following earlier work \cite{Craig:2006kx,Fischler:2006xh,Anguelova:2007at}, we have studied in great detail the static phase structure of the ISS-KKLT model when thermal corrections are considered.

We are now able to give the complete picture of its thermal evolution. At very high temperature, $\Theta \gg \Theta_c \,$, the ISS fields are at the origin because this is the point where the entropy is maximised. At these temperatures, the modulus $T$ is already stabilised (Fig. \ref{modulusplot}). Once it lies in its minimum, we can consider it to be static and neglect its quantum corrections to the ISS fields. Then, as the Universe cools down, the ISS fields start being driven away from the origin (Fig. \ref{mesonsplot}), but they are still very close to it when the temperature hits its critical value $\Theta_c \sim 10^{-7} M_P$. A second order phase transition takes place towards the would-be metastable vacua which at this stage are the true vacua of the theory. At a lower temperature, the supersymmetry preserving vacua form. They are separated from the origin by a barrier. Therefore, even if one enhances the dynamical scale $\Lambda_m$ in such a way that these vacua form first, it would consist of a first order phase transition and would thus be thermally disfavoured. For our parameters, however, the non-supersymmetric vacua form first. At a temperature $\Theta_{\rm deg} \sim 10^{-8} M_P$, the supersymmetric vacua become the global vacua of the model and from that moment on, the ISS fields can tunnel from the metastable vacua to the supersymmetric ones.

Even though we tried to give a complete and quantitative study of the model, there are still challenges that deserve further attention. First of all, we showed that, if the visible sector lives on $D3$ branes, the modulus $T$ is not destabilised by finite temperature corrections coming from the ISS sector. This assumes that the sector responsible for the stabilisation of $T$ is out of thermal equilibrium. Another limiting point that we have not treated is the dynamical evolution of the system, especially in the modulus sector. Indeed, the potential generated for a modulus is generically so steep that it seems very unlikely that the field will actually end up in the minimum, and not overshoot the barrier towards the runaway minimum\footnote{I thank Z. Lalak and S. Pokorski for bringing my attention on this problem.} (this effect is known as the Brustein-Steinhardt problem \cite{Brustein:1992nk,Dine:1999tp}). Both issues have been recently addressed in \cite{Barreiro:2007hb}. Based on the conclusions of \cite{Buchmuller:2004xr}, the authors have studied the conditions under which a stabilising sector (in their case, a SUSY-QCD) in thermal equilibrium can lead to a destabilisation of the modulus at some temperature. They developped the whole set of dynamical equations when the stabilising sector is included in the thermal fluid, and constrained the initial conditions for the rolling modulus to reach its minimum. Their conclusion is that there is a region of initial conditions which lead to a stabilisation of the modulus. The allowed region is slightly reduced compared to the case where temperature corrections are not considered, but this is not a dramatic effect. We believe that these conclusions can be applied to our case - actually the authors of \cite{Barreiro:2007hb} do study the KKLT setup - knowing that on the other hand we have showed that the temperature contribution coming from the ISS sector does not destabilise $T$. However, we think that a closer evaluation of the dynamics of our model needs to be done. In particular, thermal fluctuations around the origin might be very important.

Another interesting direction is inflation. It has been a big challenge for quite a while to combine inflation with string-inspired supergravity models : see for example \cite{Kallosh:2006dv,Brax:2007fe} and \cite{Burgess:2007pz} for a review. Here, the coupling of the ISS-flaton \cite{Craig:2008tv} to supergravity as in the ISS-KKLT setup could be of particular interest \cite{work}.


\section*{Acknowledgments}


I very warmly thank W. Buchm\"uller, E. Dudas, M. Endo, Y. Mambrini, M. Postma and A. Romagnoni for enlightening discussions, support and proofreading during the completion of this article. I also thank Z. Lalak and S. Pokorski for many discussions on the dynamics of the modulus. Part of this work was done when I was a PhD student at the LPT, Universit\'e Orsay-Paris XI and at the CPHT, \'Ecole Polytechnique, France.


\appendix

\section{Expression of the displacement of the mesons\label{appendixorigin}}

In this appendix, we derive the displacement $\Phi$ of the mesons in terms of the modulus $T$ and the temperature $\Theta$ as given in (\ref{Phigeneral}). As already sketched in Section \ref{displacedorigin}, we have to solve the equation
\be
\left . \frac{\dd V_0}{\dd \Phi} + \frac{\Theta^2}{24}\, \frac{\dd}{\dd \Phi} \, \ccl \tr \,\mathcal{M}_{f}^2 + \tr \,\mathcal{M}_{s}^2 \, \ccr \ \ar_{\ffi = \ffitil = 0} \ = \ 0 \quad ,
\label{dv/dphiappendix}
\ee
where $V_0$, $\tr \, \mathcal{M}_f^2$ and $\tr \, \mathcal{M}_s^2$ were respectively defined in (\ref{V0sugra}), (\ref{TrMfsugra}) and (\ref{TrMssugra}).

Keeping the linear order in $\Phi$, it is easy to show that the tree-level (and thus temperature independent) contribution to the displacement is
\begin{eqnarray}
\langle \, \dd_{\Phi} V_0 \, \rangle &=& \langle \, e^{K_1}\ccl  \Phib \pl K^{T \Tb} \rl D_T W_1 \Wb_{\Tb} + W_T K_{\Tb} \Wb_1 \rr + \al h^2 \mu^4 \ar N_f + \al W_1 \ar^2  \pr \right. \nonumber \\
&\quad& \quad \quad - \left. h \mu^2 \unit_{N_f} \pl K^{T \Tb} K_T \Wb_{\Tb} + \Wb_1 - 2 h \mu^{* \, 2} \tr \, \Phib \pr \ccr \, \rangle \quad ,
\label{dphiV0appendix}
\end{eqnarray}

We now turn to the fermion mass matrix and compute $\dd_{\Phi} \ccl \tr \,\mathcal{M}_{f}^2 \ccr$. The first term in (\ref{TrMfsugra}) gives the following contribution at the linear order
\begin{eqnarray}
&& \langle \, \dd_{\Phi} \pl \, e^G K^{i \bar k} K^{j \bar l} \rl G_{ij} + G_i G_j \rr  \rl G_{\bar k \bar l} + G_{\bar k} G_{\bar l} \rr \, \pr \rangle \nonumber \\
&=& \langle \, 2 e^{K_1} \ccl \ \Phib \pl h^2 \rl N + \al \mu^4 \ar N_f \rr \, \pr + \unit_{N_f} \, \al h^2 \mu^4 \ar \tr \, \Phib \, \ccr \rangle \quad .
\label{termijkl}
\end{eqnarray}

The last term is $2 e^G$ which simply gives
\be
\langle \, \dd_{\Phi} \rl -2 e^G \rr \rangle = - \langle \, 2 e^{K_1} \ccl \ \Phib \, \al W_1 \ar^2 - h \mu^2 \unit_{N_f} \rl \Wb_1 - h \mu^{* \, 2} \tr \, \Phib \, \rr \ccr \, \rangle \quad .
\label{term-2}
\ee

All together, (\ref{termijkl}) and (\ref{term-2}) give the contribution $\dd_{\Phi} \ccl \tr \, \mathcal{M}_f^2 \, \ccr$ in equation (\ref{dv/dphiappendix}).

The trace of the scalar mass matrix squared is given in (\ref{TrMssugra}) and needs the same treatment as before :

\be
\langle \, \dd_{\Phi} \rl \, 2 K^{i \bar j} \frac{\dd^2 V_0}{\dd \chi^i \dd \chib^{\bar j}} \, \rr \, \rangle \quad . \nonumber
\ee
However, with some patience, one can get the following result for this contribution
\begin{eqnarray}
&&\langle \, 2 e^{K_1} \Big[ \, \Phib \Big\{ \rl 4 + 2 N_f (N_f + 2N ) \rr \cdot \al h^2 \mu^4 \ar N_f + 2 h^2 N    \nonumber \\
&&\quad \quad \quad \quad \quad \quad \left. + \rl 1 + 2 N_f (N_f + 2N ) \rr \rl K^{T \, \Tb} D_T W_1 D_{\Tb} \Wb_1 - \al W_1 \ar^2 \rr \pr \nonumber \\
&&\quad \quad  - h \mu^2 \unit_{N_f} \Big\{ - 6 \rl 1 + N_f (N_f + 2N ) \rr h \mu^{* \, 2} \tr \, \Phib  \label{termKij} \\
&&\quad \quad \quad \quad \quad \quad \left. \left. + \rl 1 + 2 N_f (N_f + 2N ) \rr \rl K^{T \, \Tb} K_T D_{\Tb} \Wb_1 - \Wb_1 \rr  \pr  \ccr \rangle \quad , \nonumber
\end{eqnarray}
where we used the fact that $K^{i \bar j} K_{i \bar j} = 2 N_f \rl N_f + 2 N \rr$, which is a trace over the ISS scalar fields.

From all these results, it is clear that the linearised solution takes the form $\Phi = \Phi_0 \unit_{N_f}$ which implies that $\unit_{N_f} \tr \, \Phib = N_f \Phib_0$.

Eventually, plugging the different contributions into (\ref{dv/dphiappendix}), the displacement of the mesons takes the form
\be
\Phi (T, \Tb, \Theta) \ = \ h \mu^2 \rl \frac{A + B \Theta^2}{C + D \Theta^2} \rr \cdot \unit_{N_f} \quad , \nonumber
\ee
with the following entries
\begin{eqnarray}
A \rl T, \Tb \rr &=& K^{T \Tb} K_T D_{\Tb} \Wb_1 - 2 \Wb_1 \quad , \label{functionsABCD} \\
B \rl T, \Tb \rr &=& \frac{1}{12} \ccl \pl 1 + 2 N_f \rl N_f + 2 N \rr \pr \rl K^{T \Tb} K_T D_{\Tb} \Wb_1 - \Wb_1 \rr - \Wb_1 \ccr \quad , \nonumber \\
C \rl T, \Tb \rr &=& K^{T \Tb} \al D_T W_1 \ar^2 - 2 \al W_1 \ar^2 + 3 \al h^2 \mu^4 \ar N_f \quad , \nonumber \\
D \rl T, \Tb \rr &=& \frac{1}{12} \ccl \pl 1 + 2 N_f \rl N_f + 2 N \rr \pr \rl K^{T \Tb} \al D_T W_1 \ar^2 - \al W_1 \ar^2 \rr  - \al W_1 \ar^2 \right. \nonumber \\
&\quad& \quad \quad \left. + \, 3 h^2 N + \al h^2 \mu^4 \ar N_f \rl 11 + 8 N_f \rl N_f + 2 N \rr \rr \ccr \quad . \nonumber
\end{eqnarray}

\newpage

\nocite{}
\bibliographystyle{unsrt}

\end{document}